\documentclass[12pt]{article}

\usepackage{amssymb}
\usepackage{amsmath}
\usepackage[xdvi,dvips]{graphicx}

\newcommand{\br}{{\bf r}}
\newcommand{\bk}{{\bf k}}
\newcommand{\prt}{\partial}
\newcommand{\bu}{{\bf U}}

\newcommand{\om}{\omega}
\newcommand{\eps}{\varepsilon}

\begin{document}

\title{Generation of linear waves in the flow of Bose-Einstein condensate
past an obstacle}

\author{Yu.G. Gladush\footnote{\tt gladush@isan.troitsk.ru},
A.M. Kamchatnov\footnote{\tt kamch@isan.troitsk.ru}\\
\small{\it Institute of Spectroscopy, Russian Academy of Sciences}\\
\small{\it Troitsk 142190, Moscow Region, Russia}}

\maketitle

\begin{abstract}
The theory of linear wave structures generated in Bose-Einstein
condensate flow past an obstacle is developed.  The shape of wave
crests and dependence of amplitude on coordinates far enough
from the obstacle are calculated. The results are in good agreement
with the results of numerical simulations obtained earlier. The
theory gives a qualitative description of experiments with
Bose-Einstein condensate flow past an obstacle after condensate's
release from a trap.
\end{abstract}

\section{Introduction}

After Bose-Einstein condensation of dilute gases was obtained
experimentally, great attention has been devoted to excitations
existing in such systems (see, e.g., \cite{ps}). For example, it is
well known that sound waves can propagate through the condensate
with repulsive interaction between atoms. Such a condensate can
occupy considerable volume inside a trap, therefore, if a wave
length is much smaller than the condensate size, one can consider
condensate ad locally homogeneous one with a uniform density $n_0$
in an undisturbed state.  As was shown by Bogoliubov~\cite{b1947},
the linear waves have the dispersion law
\begin{equation}\label{1-1}
    \om(k)=\sqrt{c_s^2k^2+\left(\frac{\hbar k^2}{2m}\right)^2},
\end{equation}
where
\begin{equation}\label{1-2}
    c_s=\sqrt{\frac{gn_0}{m}}
\end{equation}
is a speed of sound in a long wave limit $k\to 0$, and $g$ denotes
the effective coupling constant
\begin{equation}\label{1-3}
    g=\frac{4\pi \hbar^2a_s}m
\end{equation}
arising due to $s$-wave scattering (with the scattering length
$a_s$) of atoms with mass $m$. The repulsion of atoms corresponds to
$g>0$ and a real speed of sound $c_s$. For small wave numbers the
dispersion law (\ref{1-1}) corresponds to the sound waves
propagating with speed $c_s$,
\begin{equation}\label{1-4}
    \om\cong c_s k,\qquad k\to 0,
\end{equation}
but for large $k$ it gives the quantum dispersion law of free
particles with mass $m$,
\begin{equation}\label{1-5}
    \om\cong \frac{\hbar k^2}{2m},\qquad k\to\infty.
\end{equation}
Transition from one limiting case to another takes place at
$k\sim\xi^{-1}$ where $\xi$ is a
characteristic parameter called a healing length
\begin{equation}\label{1-6}
    \xi=\frac{\hbar}{\sqrt{2mgn_0}}=\frac{\hbar}{\sqrt{2}mc_s}.
\end{equation}

If the amplitude is not small and the condensate is not homogeneous, in
the mean-field approach dynamics of the condensate is described by
the Gross-Pitaevskii equation \cite{ps}
\begin{equation}\label{2-1}
   i\hbar\frac{\prt \psi}{\prt t}=-\frac{\hbar^2}{2m}\Delta\psi+
   V(\mathbf{r})\psi+g|\psi|^2\psi,
\end{equation}
where $\psi(\br)$ is a ``condensate wave function'', and $V(\br)$ is an
external potential (for example, it might be the confining potential
of a trap). In the region where condensate is homogeneous,
potential $V$ is constant; hence, it can be excluded from the
equation by multiplication of $\psi$-function by an inessential
phase factor. Linearizing equation (\ref{2-1}) with
respect to a state with a constant density $n_0=|\psi|^2$ yields
solutions in the form of plane waves with the dispersion law
(\ref{1-1}). But appearance of the nonlinear term in (\ref{2-1})
leads to new kinds of nonlinear excitations---vortices and dark
solitons. Linear sound waves of the described origin were observed
in the experiment~\cite{andrews97} and dark solitons
in~\cite{burger,densch}. If we consider evolution of the condensate
with effectively one spatial dimension and the initial disturbance of
density is large enough, then the pulse of rarefication will decay
into a series of dark solitons~\cite{bk}. But if the initial
disturbance of density is positive, it will lead to wave breaking of
the pulse with formation of the dispersive shock
wave~\cite{kgk,damski}, as it has been found experimentally
in~\cite{simula,hoefer}.

Linear waves, dark solitons and shock waves referred above can be
described in the first approximation as waves depending only on time
and one space coordinate. But situation becomes completely different
if the wave behavior depends on two space coordinates. In
particular, the experiment~\cite{cornell05} (see also \cite{caruso})
has been made where condensate released from a trap interacted
during its expansion with a small obstacle generated by a laser
beam. As was shown experimentally, the flow past an obstacle
generates wave structures in the form of stationary ``moustaches''
attached to the obstacle. Theoretical analysis~\cite{egk1,egk2}
shows that the wave structure generated in the flow can be
subdivided into two regions separated by the Mach cone corresponding
to the sound velocity $c_s$ in the long wave limit, i.e. by lines
drawn at the angle $\theta$ to the direction of the flow so that
\begin{equation}\label{3-1}
    \sin\theta=\frac{c_s}{u_0}=\frac1M,
\end{equation}
where $M$ is the Mach number. As was shown in~\cite{egk1,egk2},
inside the Mach cone dark solitons are generated. Their slope
depends on their depth---the most shallow solitons lie near the Mach
cone and the deepest ones are located near the flow axis going
through the obstacle. In the paper~\cite{egk1} the exact stationary
solution of the Gross-Pitaevskii equation was found and it was shown
by numerical calculation that this solution describes dark solitons
generated inside the Mach cone. Linear waves are characterized by
the dispersion law (\ref{1-1}), where $k$ means the module of the
wave number. This dispersion law shows that both phase and group
velocities are greater than the sound velocity $c_s$ in a long
wavelength limit. It means that linear wave structure formed by a
wave packet can only occupy the region outside the Mach cone defined
by the equation~(\ref{3-1}). For the first time these waves
described by the Gross-Pitaevskii equation were observed in the
numerical experiment~\cite{WMCA}. In the paper~\cite{caruso}
analogous numerical simulation was carried out for description of
the structures observed in the experiments with the condensate flow
past obstacle. Some of the elementary characteristics of these
structures were derived from the dispersion law~(\ref{1-1}). In the
paper~\cite{gegk} it was noticed that these waves have a physical
origin analogous to Kelvin's ship waves~\cite{kelvin} generated by
a ship moving in a still water. It means that similar computation
methods~\cite{whitham,johnson} can be used in the case of
Bose-Einstein condensate. In~\cite{gegk} such a theory was developed
and shape of waves crests was obtained. It was shown that this
theory agrees very well with numerical simulations. But it cannot
give the dependence of the wave amplitudes on space coordinates.
The aim of this paper is to develop a complete theory which
can give not only geometrical but also dynamical properties of the
wave structures generated outside the Mach cone by the condensate
flow past obstacle.

\section{The theory of stationary wave structures generated by condensate
flow past obstacle}

As a basis of the theory, we take Kelvin's method in the form
presented in~\cite{whitham}. According to this theory, the
wave structures result from interference of linear waves
emitted by the source at all previous moments of time. We take a
source (obstacle) at rest, in agreement with the
experiment~\cite{cornell05}, and suppose that condensate moves with a
constant speed $u_0$ in the direction of the $x$ axis. Since a
stationary flow corresponds to the potential $V(\br)$ turned on in
the infinite past, one can consider this physical condition by the
assumption that the potential rises in time according to the law
$V(\br,t)=\exp(\eps t)V(\br)$. Afterwards one should take the limit
$\eps\longrightarrow0$.

It is convenient to transform the Gross-Pitaevskii
equation~$\psi(\br,t)$ to a ``hydrodynamic'' form by means of the
substitution
\begin{equation}\label{4-1}
   \psi(\br,t)=\sqrt{n(\br,t)}\exp\left(\frac{i}{\hbar}
   \int^{\br}{\bu}(\br',t)d\br'  \right ),
\end{equation}
where $n(\br,t)$ is a local density of atoms in condensate,
${\bu}(\br,t)$ is a velocity field of condensate flow. To simplify
the notation, let us introduce  non-dimensional variables
\begin{equation}\label{5-1}
    \tilde{\br}=\frac{\br}{\sqrt{2}\xi},\quad \tilde{t}=\frac{c_s}{\sqrt{2}\xi}t,
    \quad \tilde{n}=\frac{n}{n_0},\quad \tilde{\bu}=\frac{\bu}{c_s},
\end{equation}
and dimensionless potential of the obstacle
\begin{equation}\label{5-2}
    \tilde{V}(\br)=\frac1{mc_s^2}V(\sqrt{2}\,\xi\tilde{\br}).
\end{equation}
where $n_0$ is a density of undisturbed condensate (infinitely far
from the obstacle). Upon substitution of (\ref{4-1}), (\ref{5-1}) and
(\ref{5-2}) into (\ref{2-1}) and separation of real and imaginary
parts, we obtain equations for the condensate density $n$ and
velocity components $\bu=(u(x,y,t), v(x,y,t))$ in a hydrodynamic form:
\begin{equation}\label{5-3}
\begin{split}
   n_t+(nu)_x+(nv)_y&=0,\\
   u_t+uu_x+vu_y+n_x+\left(\frac{n_x^2+n_y^2}{8n^2}-
   \frac{n_{xx}+n_{yy}}{4n}\right)_x&=-V_x e^{\varepsilon t},\\
   v_t+uv_x+vv_y+n_y+\left(\frac{n_x^2+n_y^2}{8n^2}-
   \frac{n_{xx}+n_{yy}}{4n}\right)_y&=-V_y e^{\varepsilon t},
   \end{split}
\end{equation}
where tildes are omitted for convenience of the notation and the
indices stand for derivatives with respect to appropriate variables.

We are interested in linear waves propagating in a uniform flow with
$n_0=1$, $u_0=M$, $v_0=0$. Therefore we introduce new variables
\begin{equation}\label{5-4}
    n=1+n_1,\quad u=M+u_1,\quad v=v_1
\end{equation}
and linearize the system (\ref{5-3}) with respect to small deviations
$n_1,\,u_1,\,v_1$ from a uniform flow. As a result we arrive at a linear
system
\begin{equation}\label{5-5}
    \begin{split}
     n_{1,t}+u_{1,x}+Mn_{1,x}+v_{1,y}&=0,\\
     u_{1,t}+Mu_{1,x}+n_{1,x}-\tfrac14(n_{1,xxx}+n_{1,xyy})&=-V_x e^{\varepsilon t},\\
     v_{1,t}+Mv_{1,x}+n_{1,y}-\tfrac14(n_{1,xxy}+n_{1,yyy})&=-V_y e^{\varepsilon t}.
    \end{split}
\end{equation}
A stationary solution corresponds to the limit $\eps\to 0$,
where all the variables depend on time as $\exp(\eps t)$.
Fourier transform for spatial variables yields the solution
\begin{equation} {\label{6-1}}
n_1=-e^{\varepsilon t}\int\!\!\int\frac{k^2 \Phi(k_x,k_y) e^{i(k_x x+k_y y)}}
{(\eps+ik_xM)^2+k^2(1+{k^2}/{4})}\frac{dk_xdk_y}{(2\pi)^2},
\end{equation}
where
\begin{equation}\label{6-2a}
    \Phi(\bk)=\int V(\br)e^{-i\bk\br}d\br.
\end{equation}

\begin{figure}[bt]
\begin{center}
\includegraphics[width=6cm,height=6cm,clip]{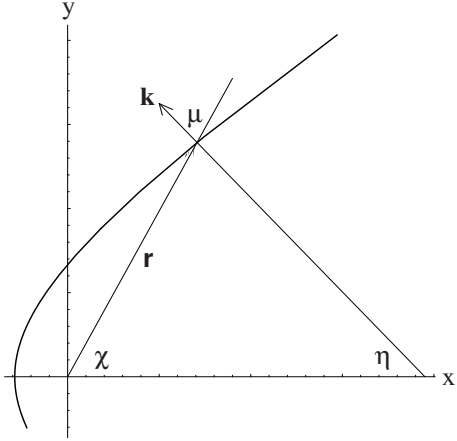}
\caption{Polar coordinates defining the radius vector $\br$ and the
wave vector $\bk$ normal to the wave front which is shown schematically by
a curved line.}
\end{center}\label{fig1}
\end{figure}
Far enough from the obstacle its potential can be considered as a
point-like one, $V(\br)=V_0\delta(\br)$; hence we have $\Phi(\bk)=V_0$.
It is convenient to introduce polar coordinates (see Fig.~1) for
vectors $\br$ and $\bk$:
\begin{equation}\label{6-2}
x=r\cos{\chi}, \quad y=r\sin{\chi}; \qquad
k_x=-k\cos{\eta},\quad k_y=k\sin{\eta}.
\end{equation}
Then simple transformation of equation (\ref{6-1}) gives
\begin{equation} \label{6-3}
n_1=\frac{V_0}{\pi^2}\,e^{\varepsilon
t}\int_{-\pi}^{\pi}\int_{0}^{\infty} \frac{ke^{-ikr\cos(\chi+\eta)}d
k d \eta}{k^2-k_0^2-i\delta\cos{\eta}},
\end{equation}
where $\delta={8M\varepsilon}/k$ is a small positive quantity and
\begin{equation}\label{6-4}
k_0=2\sqrt{M^2\cos^2{\eta}-1}.
\end{equation}
One can divide the integration interval over $\eta$ into two parts:
$\int d\eta=\int_{-\pi/2}^{\pi/2} d\eta+\int_{\pi/2}^{3\pi/2} d\eta$.
After replacement $\eta'=\eta-\pi$ in the second integral, one
can notice that the integrand function converts to its complex
conjugate one. Hence we can write the integral (\ref{6-3}) in the form
\begin{equation} \label{6-5}
n_1=\frac{2V_0}{\pi^2}e^{\varepsilon
t}\mathrm{Re}\int_{-\pi/2}^{\pi/2}\int_{0}^{\infty}
\frac{ke^{-ikr\cos(\chi+\eta)}d k d
\eta}{k^2-k_0^2-i\delta\cos{\eta}}.
\end{equation}
Integration over $k$ should be taken along the positive axis. The
pole
\begin{equation}\label{6-6}
k=\sqrt{k_0^2+i\delta\cos\eta}\cong
k_0+i\frac{\delta\cos\eta}{2k_0},
\end{equation}
lies in vicinity of a real axis, so we can expect that it gives
the main contribution into the integral.
This can be proved by the following argumentation. We add to the
integration contour along real positive $k$ axis also integration along the
positive or negative imaginary axis and a quarter of a circle with
infinite radius to make the contour closed.  If $\cos(\chi+\eta)<0$,
we use contour along the boundary of the first quadrant of complex $k$--plane,
on the other hand, if $\cos(\chi+\eta)>0$, we use the contour along
the boundary of the fourth
quadrant. In both cases integration along a quarter of a circle
vanishes when radius $|k|$ of the circle tends to infinity.
Thus, we have no contribution of this part of the integration
contour. The pole (\ref{6-6}) gets inside the integration contour
only if one take the integral along the boundary of the first quadrant
what corresponds to the case $\cos(\chi+\eta)<0$. In this case the
result of integration along the real positive axis $k$ is equal to
contribution of the residue and integration along the positive
imaginary axis. If we have $\cos(\chi+\eta)>0$, this integral is
equal to the integral along the negative imaginary axis only. In both
cases an estimation of integral along the imaginary axis gives
\begin{equation}\label{7-1}
\int_{0}^{\infty}\frac{k e^{-kr\cos(\chi+\eta)}d
k}{k^2+k_0^2}\propto\frac1{r^2},
\end{equation}
which means that this contribution decreases as $r^{-2}$ with distance $r$
from the obstacle. As we shall see below, the
contribution of the pole decreases as $r^{-1/2}$. It means
that far enough from the obstacle we can neglect the contribution of
integration along the imaginary axis.  Thus, a noticeable wave
structure appears in the region where $\cos(\chi+\eta)<0$, i.e. the
angle $\mu=\pi-(\chi+\eta)$ between vectors $\br$ and $\bk$ should
be acute. The contribution of the pole (\ref{6-6}) is equal to
\begin{equation} \label{7-2}
n_1=-\frac{2V_0}{\pi}\,\mathrm{Im}\int_{-\pi/2}^{\pi/2}e^{-ikr\cos(\chi+\eta)}d\eta,
\end{equation}
where $k$ is determined by the equation (\ref{6-4}) (index ``0''
is dropped here).

Far from the obstacle where the phase $\bk\br=rs$ is large enough, we have
large values of
\begin{equation}\label{7-3}
    s(\eta)=k(\eta)\cos(\chi+\eta),
\end{equation}
and integral (\ref{7-2}) can be estimated by a standard method of
stationary phase. Condition $\prt s/\prt\eta=0$ gives the equation
for the point of stationary phase and it can be easily transformed to
the form
\begin{equation}\label{7-4}
    \tan\mu=\frac{2M^2}{k^2}\sin2\eta
\end{equation}
or, with the help of equation for $\mu$, it gives the expression for
$\chi$,
\begin{equation}\label{7-5}
   \tan\chi=\frac{(1+k^2/2)\tan\eta}{M^2-(1+k^2/2)}.
\end{equation}
Taking into account equation (\ref{6-4}), we find
\begin{equation} \label{7-6}
\cos{\mu}=\frac{k^2}{2[(M^2-2)k^2+4(M^2-1)]^{1/2}}.
\end{equation}
With account of (\ref{7-4}), we get the expression for a second derivative of the phase
\begin{equation} \label{7-7}
\frac{\prt^2s}{\prt\eta^2}=8\frac{\cos{\mu}}{k^3}[(M^2-2)k^2+6(M^2-1)].
\end{equation}
As a result, the expression for the condensate density (\ref{7-2}) takes
the form
\begin{equation}\label{7-8}
n_1=V_0\sqrt{\frac{2k}{\pi
r}}\frac{[(M^2-2)k^2+4(M^2-1)]^{1/4}}{[(M^2-2)k^2+6(M^2-1)]^{1/2}}
\cos\left(kr\cos{\mu}-\frac{\pi}4\right).
\end{equation}

First of all, let us check that the results obtained above
agree with the theory developed in~\cite{gegk}. Since the wave
crest corresponds to the constant value of the phase $\phi=kr\cos\mu,$
we can find the equation for the absolute value of the radius-vector
\begin{equation}\label{8-1}
    r=\frac{\phi}{k\cos\mu}=\frac{2\phi}{k^3}[(M^2-2)k^2+4(M^2-1)]^{1/2}.
\end{equation}
Calculating $\cos\chi$ and $\sin\chi$ with the help of
equation~(\ref{7-5}), we obtain equations for lines of stationary
phase coinciding with those found by another method in~\cite{gegk}:
\begin{equation}\label{8-2}
    \begin{split}
    &x=r\cos\chi=\frac{4\phi}{k^3}\cos\eta(1-M^2\cos2\eta),\\
    &y=r\sin\chi=\frac{4\phi}{k^3}\sin\eta(2M^2\cos^2\eta-1).
    \end{split}
\end{equation}
Since the wave vector $k$ is determined as a function of $\eta$ by
equation~(\ref{6-4}), equations~(\ref{8-2}) give the wave crests
shape in a parametric form where parameter $\eta$ varies in the
interval
\begin{equation}\label{8-3}
    -\arccos(1/M)\leq\eta\leq\arccos(1/M).
\end{equation}

For small $\eta$ we have
\begin{equation}\label{8-4}
     \begin{split}
    &x\cong -\frac{\phi}{2\sqrt{M^2-1}}+\frac{(2M^2-1)\phi}{4(M^2-1)^{3/2}}\eta^2,\\
    &y\cong \frac{(2M^2-1)\phi}{2(M^2-1)^{3/2}}\eta,
    \end{split}
\end{equation}
that is the lines of stationary phase take here a parabolic form
\begin{equation}\label{8-5}
    x(y)\cong -\frac{\phi}{2\sqrt{M^2-1}}+\frac{(M^2-1)^{3/2}}{(2M^2-1)\phi}y^2.
\end{equation}

The limiting values  $\eta=\pm\arccos{(1/M)}$ correspond to the
lines
\begin{equation}\label{8-6}
    \frac{x}y=\pm\sqrt{M^2-1},
\end{equation}
i.e. far from the obstacle wave crest curves approach to the straight lines parallel to
those forming the Mach cone~(\ref{3-1}).

An example of the wave structure determined by
equation~(\ref{7-8}) with $M=4$ is shown on Fig.~2.

\begin{figure}[bt]
\begin{center}
\includegraphics[width=12cm,height=8cm,clip]{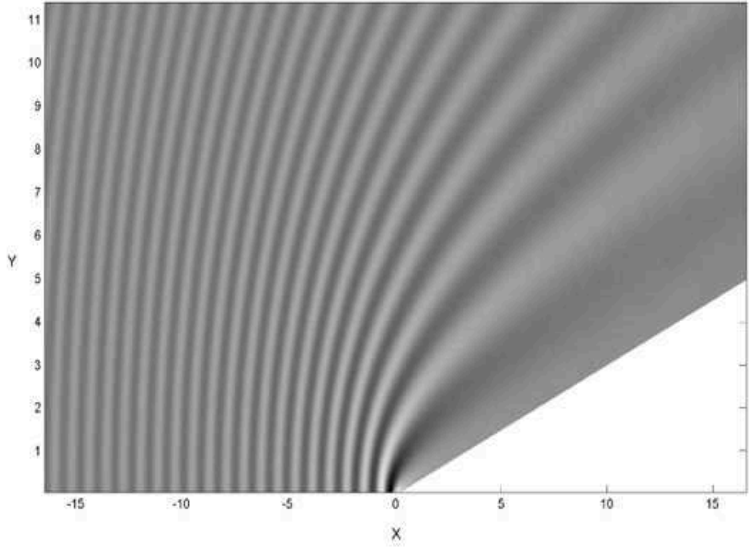}
\caption{Wave structure calculated according to equation (\ref{7-8}) with
$M=4$; it is generated outside the Mach cone and only its upper part
with $y>0$ is shown.}
\end{center}\label{fig2}
\end{figure}

If $M\gg1$, there exists a region where
$M\cos\eta\gg1$; all the equations get here a simple form. In particular,
we have $\chi\cong\pi-2\eta$, $\mu\cong\eta\cong({\pi}-{\chi})/2$,
so that
\begin{equation}\label{8-7}
    k\cong2M\sin{\frac{\chi}2}\gg 1
\end{equation}
and as a result the expression for condensate density~(\ref{7-8})
takes the form
\begin{equation}\label{8-8}
n_1\cong V_0\sqrt{\frac2{\pi
Mr}}\cos\left(2Mr\sin^2{\frac{\chi}2}-\frac{\pi}4\right).
\end{equation}
The expressions (\ref{8-2}) take here a parabolic form
\begin{equation}
x=-\frac{\phi}{2M}+\frac{M}{2\phi}y^2,
\end{equation}
Naturally, this result can be obtained from (\ref{8-3}) in the
limiting case $M\gg1$.

In the region in front of the obstacle where $y=0,\,x<0$, the
perturbations of the condensate density take the simplest form. Here
we have
\begin{equation}\label{9-1}
    k=2\sqrt{M^2-1},
\end{equation}
i.e. the wave length $\lambda=2\pi/k$ is constant and
\begin{equation}\label{9-2}
    n_1=2V_0\sqrt{\frac{(M^2-1)^{1/2}}{\pi(2M^2+1)r}}
    \cos\left[-2\sqrt{M^2-1}\,x-\frac{\pi}4\right],\quad y=0,\quad x<0.
\end{equation}
The plot illustrating this dependence is shown in Fig.~3. All the
results obtained above are in a good agreement with a numerical
simulations carried out in~\cite{gegk}.
\begin{figure}[bt]
\begin{center}
\includegraphics[width=12cm,height=8cm,clip]{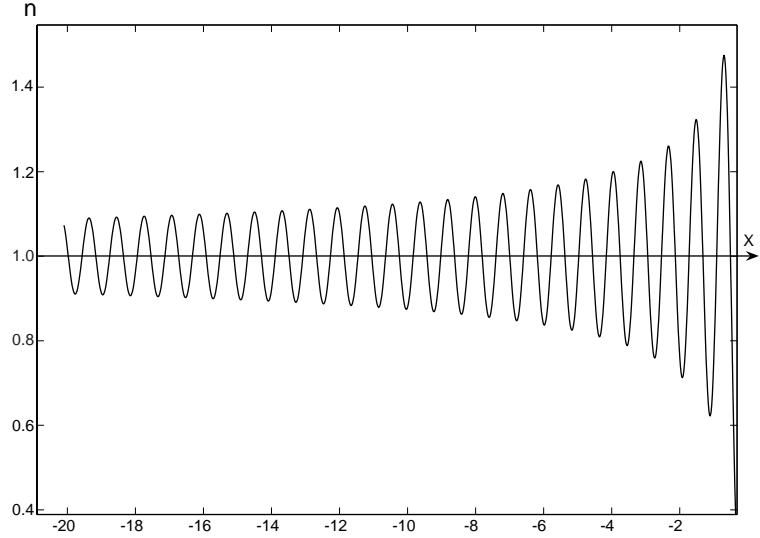}
\caption{Oscillations of the condensate density in a stationary wave in
front of the obstacle: $x<0$, $y=0$. The condensate flow
corresponds to the Mach number $M=4$.}
\end{center}\label{fig3}
\end{figure}

\section{Conclusions}
The theory developed here is valid as long as the amplitude of the wave
structure is much smaller than unity (in dimensionless variables),
i.e.
\begin{equation}\label{9-3}
    r\gg V_0^2/M.
\end{equation}
Far enough from the obstacle this condition is always fulfilled. But
for comparison of our theory with experiment it is also necessary that
the condensate flow be uniform in the region under consideration.  The estimates
made in~\cite{egk1} show that one can satisfy this condition for a
free flow of condensate released from the trap by locating the
obstacle far enough from the center of the trap and sufficiently long time
after condensate's release. As far as we can see,
this conditions were not fulfilled in the experiment \cite{cornell05,caruso}.
In particular, the obstacle was situated too close to the
center of the flow and its size was too large, so that the shadow region
appeared behind it which was not filled by the condensate. We believe that for
this reason no dark solitons predicted in \cite{egk1} were observed
experimentally. Nevertheless the wave structure in \cite{cornell05,caruso}
looks similar to one shown in Fig.~2. Thus we believe that the theory
developed above gives at least qualitative description of the wave
structures observed experimentally.

\subsection*{Acknowledgements}

We are grateful to G.A.~El and A.~Gammal  for useful discussions. This
work was supported by RFBR (grant 05-02-17351).

\end{document}